\RequirePackage{ifpdf}

\RequirePackage{fix-cm}

\documentclass[%
floatfix,
showkeys,
nofootinbib, %
superscriptaddress, %
]{revtex4-1}

\ifpdf
  \usepackage{cmap}
\fi
\usepackage{textcomp}

\usepackage[T1,T2A]{fontenc}

\usepackage[utf8]{inputenx}
\input{ix-utf8enc.dfu}

\usepackage[german,russian,english]{babel}

\usepackage{amsmath}
\usepackage{amssymb}

\usepackage{mathtools}
\mathtoolsset{
showonlyrefs,
mathic = true
}

\allowdisplaybreaks

\usepackage{hyperref}
\hypersetup{backref,
 colorlinks=false}
\hypersetup{pdfborder=0 0 0}

\usepackage{microtype}
\UseMicrotypeSet[protrusion]{alltext}

\usepackage{graphicx}

\usepackage[scanall]{psfrag}

\usepackage{listings}
\usepackage{listingsutf8}
\usepackage[frozencache]{minted}

\usepackage{fixltx2e}

\usepackage{nicefrac}

\makeatletter
\def\ps@pprintTitle{%
     \let\@oddhead\@empty
     \let\@evenhead\@empty
     \let\@oddfoot\@empty
     \let\@evenfoot\@oddfoot}
\makeatother

\usepackage{physics}

\usepackage{upgreek}
\usepackage{tensor}

\DeclareMathOperator{\ldist}{d}

\DeclareMathOperator{\adist}{\delta}

\makeatletter
\@ifpackageloaded{upgreek}{%

}{%

}
\makeatother

\begin{document}

\graphicspath{{image/red-model-approach/en/}{image/red-model-approach/}{image/}}

\title{Practical application of the multi-model approach in the study of complex systems}

\author{Anna V. Korolkova}
\email{korolkova-av@rudn.ru}
\affiliation{Department of Applied Probability and Informatics,\\
  Peoples' Friendship University of Russia (RUDN University),\\
  6 Miklukho-Maklaya St, Moscow, 117198, Russian Federation}

\author{Dmitry S. Kulyabov}
\email{kulyabov-ds@rudn.ru}
\affiliation{Department of Applied Probability and Informatics,\\
  Peoples' Friendship University of Russia (RUDN University),\\
  6 Miklukho-Maklaya St, Moscow, 117198, Russian Federation}
\affiliation{Laboratory of Information Technologies\\
  Joint Institute for Nuclear Research\\
  6 Joliot-Curie, Dubna, Moscow region, 141980, Russian Federation}

\author{Michal Hnati\v{c}}
\email{hnatic@saske.sk}
\affiliation{Department of Theoretical Physics, SAS, Institute of
  Experimental Physics,\\
Watsonova 47, 040 01 Ko\v{s}ice, Slovak Republic}
\affiliation{Pavol Jozef \v{S}af\'{a}rik University in Ko\v{s}ice (UPJ\v{S}),\\
  \v{S}rob\'{a}rova 2, 041 80 Ko\v{s}ice, Slovak Republic}
\affiliation{Bogoliubov Laboratory of Theoretical Physics\\
  Joint Institute for Nuclear Research\\
  6 Joliot-Curie, Dubna, Moscow region, 141980, Russian Federation}

\begin{abstract}
Different kinds of models are used to study various natural and
technical phenomena. Usually, the researcher is limited to using a
certain kind of model approach, not using others (or even not
realizing the existence of other model approaches). The authors
believe that a complete study of a certain phenomenon should cover
several model approaches. The paper describes several model
approaches which we used in the study of the random early detection algorithm for
active queue management. Both the model approaches
themselves and their implementation and the results obtained are
described.
\end{abstract}


\maketitle

\section{Introduction}
\label{sec:intro}

  Scientific research is easy to start but difficult to complete. Our
  study of the Random Early Detection (RED) algorithm stood out from
  the study of approaches and mechanisms of traffic control in data
  transmission networks. But the further we went, the less satisfied
  we were with the results. The originally constructed mathematical
  model seemed to us somewhat artificial and non-extensible. To build
  a more natural mathematical model from first principles, we have
  developed a method of stochastization of one-step processes. To
  verify the mathematical model, we have built physical and simulation
  models. To conduct optimization studies, we began to build a
  surrogate model for the RED algorithm. At the last we came to an
  understanding that all of our models form some kind of emergent
  structure, with the help of which we can investigate various
  phenomena --- stochastic and statistical systems in particular.

  In this paper, we try to present our understanding of the
  multi-model approach to modeling.

\section{Model approaches}
\label{sec:model-approach}

\begin{figure}[b]
  \centering
  \includegraphics[width=0.6\linewidth]{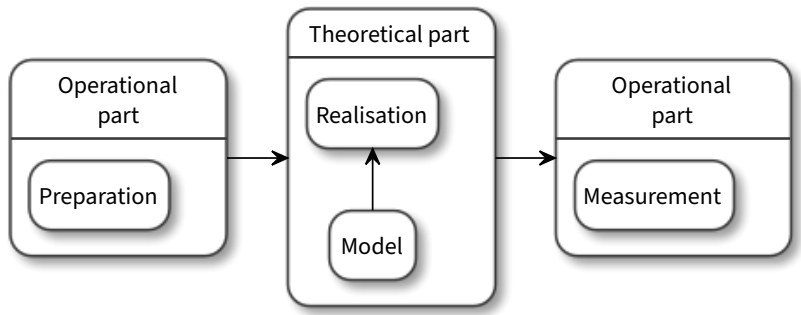}
  \caption{Generic structure of the model approach}
  \label{fig:model-generic}
\end{figure}

  Modeling as a discipline encompasses different types of model
  approaches. From our point of view, these approaches can be
  schematically described in a unified manner (see
  Fig.~\ref{fig:model-generic}). In this case, the research structure
  consists of operational and theoretical parts. The operational parts
  are represented by procedures of the system preparation and measurement. 
  It is also common to describe the operational part as
  input and output data.

  The theoretical part consists of two layers: a model layer and an
  implementation layer. The implementation layer describes the
  specific structure of the evolution of the system.  Depending on the
  type of implementation, different types of models can be obtained: a
  mathematical model (implementation~--- mathematical expressions), a
  simulation model (implementation~--- an algorithm), a physical model
  (implementation~--- an analog system), a surrogate model
  (implementation~--- approximation of behavior).  Each type of model
  has its area of applicability, its advantages and disadvantages. The
  use of the entire range of models allows the most in-depth and
  comprehensive study of the modeled system.

\section{RED active queue management algorithm}
\label{sec:red-aqm}

  Random Early Detection (RED) algorithm is at the heart of several mechanisms
  to prevent and control congestion in router queues. Its main purpose
  is to smooth out temporary bursts of traffic and prevent prolonged
  network congestion by notifying traffic sources about the need to
  reduce the intensity of information transmission.

  The operation of a module implementing a RED-type algorithm 
  can be schematically represented as follows.

  When a packet of transmitted data enters the system, it enters the
  reset module. The decision to remove the package is made based on
  the value of the $p(\Hat{q})$ function received from the
  control unit. The function $p(\Hat{q})$ depends on the
  exponentially weighted moving average of the queue length
  $\Hat{q}$, also calculated by the ontrol unit based on the
  current value of the queue length $q$.

  The classic RED algorithm is discussed in detail
  in~\cite{floyd:1993:red}. Only the formulas for calculating the
  reset function $p(\Hat{q})$ and the exponentially weighted moving
  average queue length~$\Hat{q}$ are given here. The use of $\Hat{q}$
  is associated with the need to smooth outliers of the instantaneous
  queue length $q$.

  To calculate $\Hat{q}$, the recurrent formula for exponentially
  weighted moving average (EWMA) is used:
\begin{equation}
\label{eq:barq}
\Hat{q}_{k+1} = (1-w_q)\Hat{q}_k +w_q q_k, \quad k=0,1,2,\ldots,
\end{equation}
  where $w_q$, $0 <w_q <1$ is the weight coefficient of the exponentially weighted
  moving average:
\begin{equation}
w_q=1-e^{-1/C},
\end{equation}
where $ C $ is the channel bandwidth (packets per second).
 
The $p(\Hat{q})$ packet discard function linearly depends on the
$\Hat{q}$, the minimum $q_{\min}$ and maximum $q_{\max}$
thresholds and the maximum discard parameter $p_{\max}$, which
sets the maximum packet discard level when $\Hat{q}$ reaches the
value $q_{\max}$, and the function $p(\Hat{q})$ is calculated as follows:
\begin{equation}
\label{eq:p:red}
p(\Hat{q}) =
\begin{dcases}
0, & 0 < \Hat{q} \leqslant  q_{\min},\\
 \frac{\Hat{q}-q_{\min}}{q_{\max}-q_{\min}} p_{\max}, & q_{\min}< \Hat{q} \leqslant  q_{\max},\\
1, & \Hat{q} >  q_{\max}.
\end{dcases}
\end{equation}

The drop function~\eqref{eq:p:red} describes the classic RED
algorithm. And the main effort in the design of new algorithms like
RED is directed at various modifications of the type of the drop
function.

Since the complete simulated system consists of interoperable TCP and
RED algorithms, it is necessary to simulate the evolution of the TCP
source as well. Since the original model was based on the TCP Reno
protocol, we simulated this particular protocol.

TCP uses a sliding window mechanism to deal with congestion. The
implementation of this mechanism depends on the specific TCP protocol
standard.

In TCP Reno the congestion control mechanism consists of the
following phases: slow start, congestion avoidance, fast transmission,
and fast recovery. The dynamics of the congestion window (CWND) size
depends on the specific phase.

TCP Reno monitors two types of packet loss:
\begin{itemize}
\item Triple Duplicate ACK (TD).  Let the $n$-th packet  not to be 
  delivered, and the subsequent packets ($n + 1$, $n + 2$, etc.)
  are delivered. For each packet delivered out of order (for
  $n + 1$, $n + 2$, etc.), the receiver sends an ACK message for
  the last undelivered ($n$-th) packet. Upon receipt of three such
  packets, the source resends the $n$-th packet. Besides, the
  window size is reduced by 2 times $cwnd \to cwnd/2$.
\item Timeout (TO). When a packet is sent, the timeout timer is
  started. Each time a confirmation is received, the timer is
  restarted. The window is then set to the initial value of the
  overload window. The first lost packet is resent. The protocol
  enters the slow start phase.
\end{itemize}

The general congestion control algorithm is of the AIMD type (Additive
Increase, Multiplicative Decrease)~--- an additive increase of the
window size and its multiplicative decrease.

\section{Mathematical model}
\label{sec:math-model}

\begin{figure}
  \centering
  \includegraphics[width=0.6\linewidth]{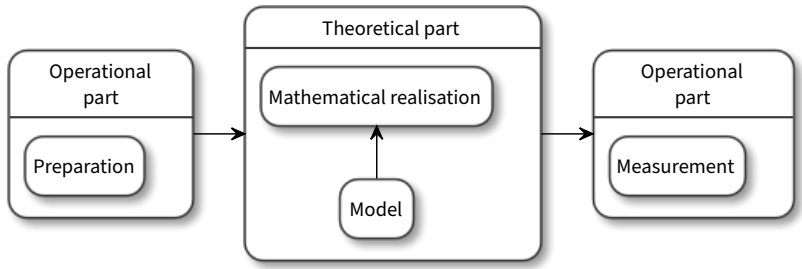}
  \caption{Generic structure of the mathematical model}
  \label{fig:model-math}
\end{figure}

The most rigorous research is usually based on a mathematical model
(see Fig.~\ref{fig:model-math}). In this case, the model layer is
realized through mathematical expressions describing the evolution of
the system.

There are several approaches to modeling RED-type algorithms. The most
famous approach is modeling using the automatic control theory
approach~\cite{misra:1999:sdu,misra:2000:fluid-based,hollot:2001:control}. To us, this approach seems somewhat artificial
and inconsistent. We prefer to do our modeling from first principles.

We have developed a method of stochastization of one-step processes,
which allows us to obtain models from first principles. Moreover, the
resulting model models are immanently
stochastic~\cite{kulyabov:2018:pcs:stochastization_computer_algebra::en,kulyabov:2016:mmcp:one-step,kulyabov:2016:ecms:one-step}.  Our
model of interaction between the TCP source and the RED algorithm is
based on these methods and is mathematically represented in the form
of stochastic differential equations with Wiener and Poisson
processes~\cite{kulyabov:2019:ecms:red-control,kulyabov:2017:iopconf:self-oscillations,kulyabov:2014:vestnik:red-sdu::en}.

  We will use the following notation. $W (t)$ is the TCP Reno window
  size, $ Q (t) $ is the queue size, $ T: = T (Q (t)) $ is the round-trip
  time (taking into account equipment delays),
  $t$ is the time, $C$ is the queue service intensity,
  $\Hat{Q}(t)$ is the Exponentially Weighted Moving-Average (EWMA)~\cite{floyd:1993:red}:
\begin{equation}
  \label{qhat}
\Hat{Q}(t)= (1-w_q)\Hat{Q}(t)+w_q Q(t),
\end{equation}
where $w_q$, $0 < w_q < 1$ is the weight coefficient.

  We used the method of stochastization of one-step processes to
  obtain the Fokker--Planck and Langevin equations for the stochastic
  processes $W (t)$ and $Q (t)$.

The kinetic equations will be:
\begin{equation}
  \label{eq:kinetik:W:Q}
  \begin{dcases}
    0 \xrightarrow{\frac{1}{W(t)}} W (t), \\
    W (t) \xrightarrow{\frac{1}{2} \frac{\dd N(t)}{\dd t}} 0, \\
    0 \xrightarrow{\frac{W(t)}{T}} Q (t),\\
    0\xrightarrow{ - C} Q(t),
  \end{dcases}
\end{equation}
where $\dd N(t)$ is the Poisson process~\cite{misra:2000:fluid-based}.

  Let us write down the Fokker--Planck equations corresponding to the
  kinetic equations~\eqref{eq:kinetik:W:Q}:
\begin{equation}
  \label{eq:FP:W:Q}
  \begin{gathered}
    \begin{multlined}
  \frac{\partial w(t)}{\partial t} = 
  - \frac{\partial}{\partial W(t)} 
  \left[ \left( \frac{1}{W(t)} - \frac{W(t)}{2} \frac{\dd N(t)}{\dd t}
    \right) w(t) \right]
  + {} \\ {} 
  +
  \frac{1}{2} \frac{\partial^2}{\partial W^2(t)} 
  \left[ \left( \frac{1}{W(t)} + \frac{W(t)}{2} \frac{\dd N(t)}{\dd t}
    \right) w(t) \right],
\end{multlined}
\\
\begin{multlined}
  \frac{\partial q(t)}{\partial t} = - \frac{\partial}{\partial
    Q(t)}
  \left[ \left( \frac{W(t)}{T} - C \right) q(t) \right]
  + {} \\ {}
  +
  \frac{1}{2}
  \frac{\partial^2}{\partial Q^2(t)} \left[ \left( \frac{W(t)}{T} -
      C \right) q(t) \right],
\end{multlined}
\end{gathered}
\end{equation}
  where $w(t)$  is the density of the random process $W(t)$,
  $q(t)$ is the density of the random process $Q(t)$.

  The Langevin equations corresponding to the
  equations~\eqref{eq:FP:W:Q} have the form:
\begin{equation}
  \label{eq:W:Q}
  \begin{dcases}
      \dd W(t) = \frac{1}{T} \dd t - \frac{W(t)}{2} \dd N(t)
      +
      \sqrt{\frac{1}{T} + \frac{W(t)}{2} \frac{\dd N(t)}{\dd t}} \dd
      V^1(t),
    \\
    \dd Q(t)  = \left( \frac{W(t)}{T} - C \right) \dd Q(t) + \sqrt{\frac{W(t)}{T}
      - C}\dd V^2(t) ,
  \end{dcases}
\end{equation}
  where $\dd V^1 (t)$ is the Wiener process corresponding to the
  random process $W(t)$, $\dd V^2 (t)$ is the Wiener process
  corresponding to the random process $Q(t)$.

  The equations~\eqref{eq:W:Q} are supplemented by the constraint
  equation (written in differential form for convenience):
\begin{equation}
  \label{eq:Qhat}
  \frac{\dd \Hat{Q}(t)}{\dd t}  = w_q C ( Q(t) - \Hat{Q}(t) ).
\end{equation}

This mathematical model can be investigated both in the form of
stochastic differential equations and by writing them down in
moments. The equation in moments is naturally easier to study.

\section{Physical model}
\label{sec:physical-model}

The resulting mathematical model should be compared with experimental
data and verified. Unfortunately, we do not have the resources to take
data from a working network or build a full-scale test bench on real
network equipment. Therefore, we tried to create a virtual
experimental installation based on the virtual machines~\cite{kulyabov:2014:icumt-2014:gns3}.
Virtual machines run images of real routers operating systems. This is
what allows us to call this model \emph{physical}.

To create the stand, the software package GNS3 (Graphical Network
Simulator)~\cite{welsh:2013:gns3} was chosen. This allows you to
simulate a virtual network of routers and virtual machines. The software package GNS3 works on
almost all platforms. It can be considered as a graphical interface for different
virtual machines. To emulate Cisco devices the Dynamips emulator is
used. Alternatively, emulators such as VirtualBox and Qemu can be
used. The latter is especially useful when it used with a KVM system which
allows a hardware processor implementation. GNS3 coordinates the
operation of various virtual machines and also provides the
researcher with a convenient interface for creation and customization of 
the required stand configuration. Also, the developed topology can be
linked to an external network to manage and control data packets.

\begin{figure}
  \centering
  \includegraphics[width=0.6\linewidth]{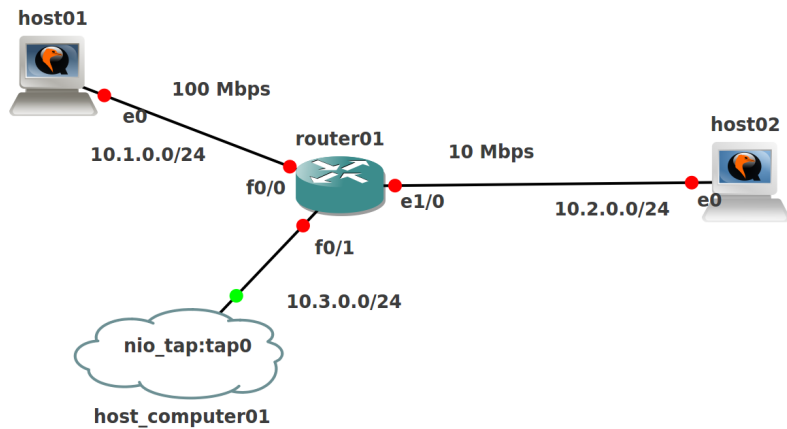}
  \caption{Virtual stand for studying the functioning of the RED algorithm. \texttt{host01} is the packet source; \texttt{host02} is the recipient.}
  \label{fig:stand}
\end{figure}

The stand consists of a Cisco router, a traffic generator, and a
receiver. D-ITG (Distributed Internet Traffic Generator) is used as a
traffic generator (see Fig.~\ref{fig:stand}). D-ITG allows us to obtain
estimates of the main indicators of the quality of service (average
packet transmission delay, delay variation (jitter), packet loss rate,
performance) with a high degree of confidence.

\section{Simulation model}
\label{sec:simulation}

\begin{figure}
  \centering
  \includegraphics[width=0.6\linewidth]{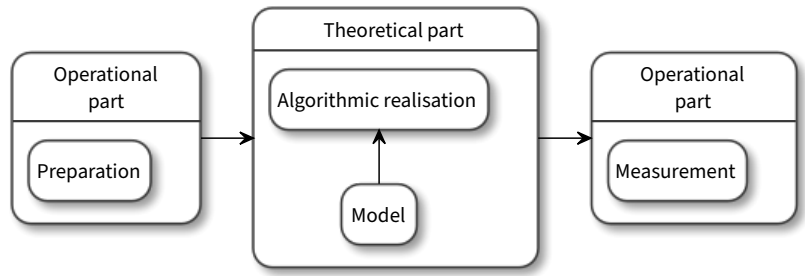}
  \caption{Generic structure of the simulation model}
  \label{fig:model-simulation}
\end{figure}

With the development of computer technology it became possible to
specify a model implementation not in the form of a mathematical
description, but in the form of some algorithm
(Fig.~\ref{fig:model-simulation}). This type of model is called
simulation models 
and the approach itself is called simulation.

The simulation model plays a dual role. The simulation model, debugged
and tested on experimental data and a physical model, can itself serve
the purposes of the mathematical model verification.
On the other hand, the simulation model makes it possible to study the
behavior of the modeled system more effectively than the mathematical
model for different variants of the input data.

\subsection{Simulation model on NS-2}
\label{sec:red:ns2}

The \texttt{ns2}~\cite{issariyakul:2012,altman:2011} package is the network
protocol simulation tool. During its existence, the functionality has
been repeatedly verified by data from field experiments. Therefore,
this package itself has become a reference modeling tool. This is
exactly the case when a simulation model is a replacement for a
physical model and a natural experiment.

\begin{figure}
  \centering
  \includegraphics[width=0.7\linewidth]{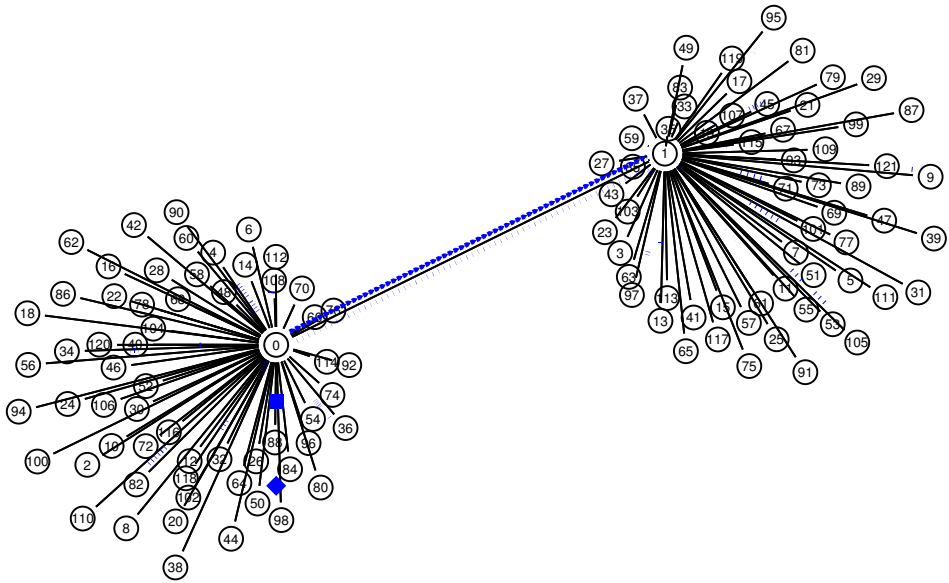}
  \caption{Visualization of the simulation. Packets drop is shown}
  \label{fig:nam-drop}
\end{figure}

The program for ns2 is written in the TCL
language~\cite{welch:book:practical-tc-tk,nadkarni:book:tcl-comprehensive}.  The simulation results can be
represented using visualization tool \texttt{nam} (see
Fig.~\ref{fig:nam-drop}).

The simulator is built on an event-driven architecture. That is, it
implements a discrete approach to modeling. On the one hand, this is a
plus, since it directly implements the TCP and RED specification (see
section~\ref{sec:red-aqm}). On the other hand, the amount of resulting
data sharply increases, which makes it difficult to carry out any
lengthy simulation experiment.

In our works, this software is used precisely for verification of 
the obtained results~\cite{kulyabov:2018:icumt:aqm,kulyabov:2019:aisc:self-oscillation}.

\subsection{Hybrid model for RED algorithm}
\label{sec:red:model_hybrid}

To study the RED algorithm we developed the prototype of the simulation
model. We wanted to avoid the resource intensiveness of discrete
modeling approaches. However, it was necessary to take into account
the discrete specifics of TCP and RED (see
section~\ref{sec:red-aqm}). Therefore, we have chosen a hybrid
(continuous--discrete) approach. The model was implemented in the
hybrid modeling language Modelica~\cite{fritzson:2003,fritzson:2011}.

Since we are building the hybrid continuous--discrete model, then to
describe each phase of TCP functioning, we will turn to the model with
continuous time. The transition between phases will be described by
discrete states.

To build a hybrid model, we need:
\begin{itemize}
\item write a dynamic model for each state;
\item replace systems with piecewise constant parameters with systems
  with variable initial conditions;
\item write the state diagram of the model (Fig.~\ref{fig:tcp:general}
  and~\ref{fig:red:general}).
\end{itemize}

The resulting diagrams are directly converted into a Modelica
program~\cite{kulyabov:2017:iop:tcp-modelica,kulyabov:2019:ceur-ws:2407:red,kulyabov:2016:ecms:red}.

\begin{figure}
\begin{minipage}[b]{0.6\linewidth}
  \centering
  \includegraphics[width=\linewidth]{tcp_general}
  \caption{TCP state diagram}
  \label{fig:tcp:general}
\end{minipage}
\hfill
\begin{minipage}[b]{0.35\linewidth}
  \centering
  \includegraphics[width=\linewidth]{red_general}
  \caption{RED state diagram}
  \label{fig:red:general}
\end{minipage}
\end{figure}

\section{Surrogate model}
\label{sec:surrogate}

\begin{figure}
  \centering
  \includegraphics[width=0.6\linewidth]{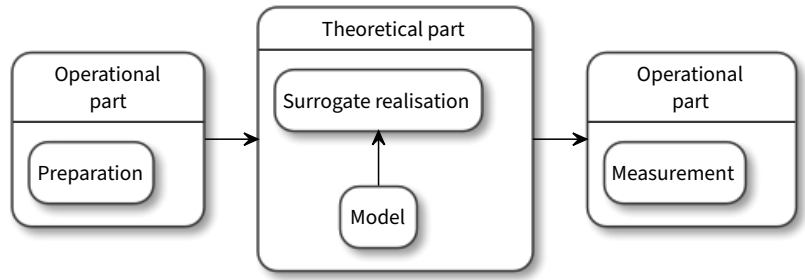}
  \caption{Generic structure of the surrogate model}
  \label{fig:model-surrogate}
\end{figure}

Most scientific and technical problems require experiments and
simulations to obtain results, to determine the limitations imposed on
the result. However, for many real-world problems simulation alone
can take minutes, hours, days. As a result, routine tasks such as
decision optimization, decision space exploration, sensitivity
analysis, and what-if analysis become impossible as they require
thousands or millions of modeling evaluations.

One way to simplify research is to build surrogate models
(approximation models, response surface models, metamodels, black box
models) (see Fig.~\ref{fig:model-surrogate}) that mimic the behavior
of the original model so closely as much as possible, while being
computationally
cheap~\cite{jin:2011:surrogate-assisted-computation}. Surrogate models
are built using a data-driven approach. The exact inner workings of
the simulation code are not supposed to be known (or even understood),
only the input---output (preparation---measurement) behavior is
important. The model is built based on modeling the response to a
limited number (sometimes quite large) of selected data points
\footnote{Note that this type of model is known to many
  researchers. When only one modeling variable is involved, the
  process of building the surrogate model is called curve fitting}.

The scientific challenge for surrogate modeling is to create a
surrogate that is as accurate as possible using as few modeling
estimates as possible. The process consists of the following main
stages~\cite{kulyabov:2019:ceur-ws:2507:deep-learning}:
\begin{itemize}
\item sample selection;
\item construction of the surrogate model and optimization of model
  parameters;
\item an estimate of the accuracy of the surrogate.
\end{itemize}

For some problems the nature of the true function is a priori
unknown, so it is unclear which surrogate model will be the most
accurate. Moreover, it is not clear how to obtain the most reliable
estimates of the accuracy of a given surrogate. In this case, the
model layer (Fig.~\ref{fig:model-surrogate}) is replaced by the
researcher's guess. In our case, the surrogate model is based on a
clearly formulated mathematical model, which allows us to obtain
clear, substantiated results of surrogate modeling.

At this moment we are developing a methodology for constructing
surrogate models for both RED type algorithms and
arbitrary stochastic one-step processes.

\section{Conclusion}
\label{sec:conclusion}

The authors tried to outline the concept of a multi-model approach to
the study of physical and technical systems using the example of the
interaction between the TCP protocol and the RED-type active queue management
algorithm. This research is in line with the research of stochastic
models in science and technology.

The multi-model approach makes it possible to increase the efficiency
of the study of the phenomenon, to consider it from different angles,
and to create effective software systems.

\begin{acknowledgments}

The publication has been prepared with the support of the ``RUDN University Program 5-100''
and funded by Russian Foundation for Basic Research (RFBR) according to the research project
No~19-01-00645.

\end{acknowledgments}

 \bibliographystyle{elsarticle-num}

\bibliography{bib/red-model-approach/cite}

\end{document}


\graphicspath{{image/red-model-approach/ru/}{image/red-model-approach/}{image/}}

\title{Практическое применение мультимодельного подхода при исследовании сложных систем}

\author{А. В. Королькова}
\email{korolkova-av@rudn.ru}
\affiliation{Кафедра прикладной информатики и теории вероятностей,\\
  Российский университет дружбы народов,\\
  117198, Москва, ул. Миклухо-Маклая, д. 6}

\author{Д. С. Кулябов}
\email{kulyabov-ds@rudn.ru}
\affiliation{Кафедра прикладной информатики и теории вероятностей,\\
  Российский университет дружбы народов,\\
  117198, Москва, ул. Миклухо-Маклая, д. 6}
\affiliation{Лаборатория информационных технологий,\\
  Объединённый институт ядерных исследований,\\
  ул. Жолио-Кюри 6, Дубна, Московская область, Россия, 141980}

\author{М. Гнатич}
\email{hnatic@saske.sk}
\affiliation{Лаборатория теоретической физики,\\
  Объединённый институт ядерных исследований,\\
  ул. Жолио-Кюри 6, Дубна, Московская область, Россия, 141980}

\begin{abstract}

Для исследования разнообразных природных и технических явлений
используется разные типы моделей. Обычно исследователь ограничивается
использованием какого-то определённого типа модельного подхода, не
используя другие (или даже не осознавая существование других модельных подходов).
Авторы считают, что полное исследование некоторого явления должно
охватывать несколько модельных подходов. В работе описываются
несколько модельных подходов, использованные нами при исследовании
модуля активного управления трафиком RED. Описываются как сами
модельные подходы, так и их реализации и полученные результаты.

\end{abstract}

  \keywords{активное управление трафиком,
    математическое моделирование, 
    имитационное моделирование,
    суррогатное моделирование,
    стохастические системы}

\maketitle

\section{Введение}
\label{sec:intro}

Научное исследование легко начать, но трудно завершить. Наше
исследование алгоритма RED выделилось из исследования подходов и
механизмов управления трафиком в сетях передачи данных. Но чем дальше
мы продвигались, тем менее нас удовлетворяли полученные
результаты. Первоначально построенная математическая модель казалась
нам несколько искусственной и нерасширяемой. Для построения более
естественной математической модели из первых принципов нами была
разработана методика стохастизации одношаговых процессов. Для
верификации математической модели нами были построены физическая и
имитационные модели. Для проведения оптимизационных исследований мы
стали строить суррогатную модель алгоритма RED. В конце концов мы
достигли понимания, что все наши модели образуют некоторую
эмерджентную структуру, с помощью которой можно исследовать
разнообразные явления. В частности, стохастические и статистические
системы.

В этой работе мы пытаемся представить наше понимание мультимодельного
подхода к моделированию.

\section{Модельные подходы}
\label{sec:model-approach}

Моделирование как дисциплина охватывает разные типы модельных
подходов. С нашей точки зрения эти подходы можно схематически описать
единым образом (см. рис.~\ref{fig:model-generic}). В данном случае
структура исследования состоит из операциональных и теоретических
частей. Операциональные части представлены процедурами приготовления
системы и измерения. Также распространено описание операциональной
части как входных и выходных данных.

Теоретическая часть состоит из двух слоёв: модельного слоя и слоя
реализации. Слой реализации описывает конкретную структуру эволюции
системы.
В зависимости от типа реализации, можно получать разные виды моделей:
математическую модель (реализация~--- математические выражения),
имитационная модель (реализация~--- алгоритм), физическая модель
(реализация~--- аналоговая система), суррогатная модель
(реализация~--- аппроксимация поведения).
Каждый тип моделей имеет свою область применимости, свои преимущества
и недостатки. Использование всего спектра моделей позволяет наиболее
глубокое и всестороннее исследование моделируемой системы.

\begin{figure}[b]
  \centering
  \includegraphics[width=0.6\linewidth]{model-generic}
  \caption{Общая структура модельного подхода}
  \label{fig:model-generic}
\end{figure}

\section{Алгоритм активного управления трафиком RED}
\label{sec:red-aqm}

Алгоритм Random Early Detection (RED) лежит в основе ряда механизмов
предотвращения и контроля перегрузок в очередях маршрутизаторов. Его
основное предназначение заключается в сглаживании временных всплесков
трафика и предупреждении длительной перегрузки сети посредством
уведомления источников трафика о необходимости снижения интенсивности
передачи информации.

Функционирование модуля, реализующего алгоритм типа RED, можно
схематично представить следующим образом.

Пакет передаваемых данных при поступлении в систему попадает в модуль
сброса. Решение о сбросе пакета принимается на основе значения функции
$p(\Hat{q})$, получаемого от управляющего модуля. Функция $p(\Hat{q})$ 
зависит от значения экспоненциально взвешенного скользящего среднего
размера длины очереди $\Hat{q}$, также вычисляемого управляющим
модулем, основываясь на текущем значении длины очереди $q$.

Классический алгоритм RED подробно рассмотрен в
работе~\cite{floyd:1993:red}.  Здесь приведены лишь формулы для
расчёта функции сброса $p(\Hat{q})$ и экспоненциально взвешенной
скользящей средней длины очереди~$\Hat{q}$.  Использование $\Hat{q}$
вызвано необходимостью сглаживания выбросов мгновенной длины очереди
$q$.

Для вычисления $\Hat{q}$ используется рекуррентная формула экспоненциально
взвешенного скользящего среднего (Exponentially Weighted
Moving-Average, EWMA):
\begin{equation}
\label{eq:barq}
\Hat{q}_{k+1} = (1-w_q)\Hat{q}_k +w_q q_k, \quad k=0,1,2,\ldots,
\end{equation}
где $w_q$, $0 < w_q < 1$~--- весовой коэффициент экспоненциально
взвешенного скользящего среднего:
\begin{equation}
w_q=1-e^{-1/C},
\end{equation}
где $C$~--- пропускная способность канала (пакетов в секунду).
 
Функция $p(\Hat{q})$ сброса пакетов линейно зависит от $\Hat{q}$,
минимального $q_{\min}$ и максимального $q_{\max}$ пороговых значений
и параметра максимального сброса $p_{\max}$, задающего максимальный
уровень сброса пакетов при достижении $\Hat{q}$ значения $q_{\max}$, и
вычисляется следующим образом:
\begin{equation}
\label{eq:p:red}
p(\Hat{q}) =
\begin{dcases}
0, & 0 < \Hat{q} \leqslant  q_{\min},\\
 \frac{\Hat{q}-q_{\min}}{q_{\max}-q_{\min}} p_{\max}, & q_{\min}< \Hat{q} \leqslant  q_{\max},\\
1, & \Hat{q} >  q_{\max}.
\end{dcases}
\end{equation}

Данная функция сброса~\eqref{eq:p:red} описывает классический алгоритм
RED. И основные усилия при конструировании новых алгоритмов типа RED
направлено на разные модификации вида функции сброса.

Поскольку полная моделируемая система состоит из взаимодействующих
протокола TCP и алгоритма RED, то необходимо моделировать также и
эволюцию источника, работающего по протоколу TCP. Поскольку
оригинальная модель базировалась на протоколе TCP Reno, то нами
моделировался именно этот протокол.

В протоколе TCP используется механизм скользящего окна для борьбы с
перегрузками. Реализация данного механизма зависит от конкретного
стандарта протокола TCP.

В TCP Reno механизм управления перегрузками состоит из следующих фаз:
медленный старт, предотвращение перегрузок, быстрая передача и быстрое
восстановление.  Динамика изменения размера окна перегрузки
(Congestion Window, CWND) зависит от конкретной фазы.

Протокол TCP Reno отслеживает два варианта потери пакетов:
\begin{itemize}
\item Тройное дублирование подтверждения (Triple Duplicate ACK,
  TD).
  Пусть $n$-й пакет не доставлен, а последующие пакеты ($n+1$,
  $n+2$ и т.д.) доставлены. Для каждого пакета, доставленного в
  нарушении очерёдности (для $n+1$, $n+2$ и т.д.) получатель отсылает
  сообщение ACK для последнего недоставленного ($n$-го) пакета. При
  получении трёх таких пакетов источник перепосылает $n$-й
  пакет. Кроме того, размер окна уменьшается в 2 раза $cwnd \to
  cwnd/2$.
\item Тайм-аут (Timeout, TO).
  При отправке пакета запускается таймер
  тайм-аута. При получении каждого подтверждения таймер
  перезапускается. Окно при этом устанавливается в начальное значение
  окна перегрузки. Первый потерянный пакет перепосылается. Протокол
  переходит в фазу медленного старта.
\end{itemize}

Общий алгоритм управления перегрузкой относится к типу AIMD (Additive
Increase, Multiplicative Decrease)~--- аддитивное увеличение размера
окна и мультипликативное его уменьшение.

\section{Математическая модель}
\label{sec:math-model}

\begin{figure}
  \centering
  \includegraphics[width=0.6\linewidth]{model-math}
  \caption{Общая структура математической модели}
  \label{fig:model-math}
\end{figure}

Наиболее строгое исследование базируется, обычно, на математической
модели (см. рис.~\ref{fig:model-math}). В этом случае модельный слой
реализуется посредством математических выражений, описывающих эволюцию
системы.

Есть несколько подходов к моделированию алгоритмов типа RED. Наиболее
известным подходом является моделирование с использование подхода
теории автоматического управления~\cite{misra:1999:sdu,misra:2000:fluid-based,hollot:2001:control}.
Нам этот подход представляется несколько искусственным и
непоследовательным. Мы предпочитаем проводить моделирование из первых принципов.

Нами разработан метод стохастизации
одношаговых процессов, который позволяет получать модели из первых
принципов. При этом полученные модели модели являются имманентно
стохастическими~\cite{kulyabov:2018:pcs:stochastization_computer_algebra::ru,kulyabov:2016:mmcp:one-step,kulyabov:2016:ecms:one-step}.
Наша модель взаимодействия источника TCP и алгоритма RED
строится на основе этих методов и математически представляется в виде
стохастических дифференциальных уравнений с винеровским и пуассоновским
процессом~\cite{kulyabov:2019:ecms:red-control,kulyabov:2017:iopconf:self-oscillations,kulyabov:2014:vestnik:red-sdu::ru}.

Будем использовать следующие обозначения.
$W(t)$ --- функция изменения размера окна TCP Reno, $Q(t)$
--- функция изменения размера очереди, $T:=T(Q(t))$~--- время двойного
оборота c учётом задержек обработки
пакетов в оборудовании,
$t$~--- время, $C$~--- интенсивность обслуживания
пакетов в очереди,
$\Hat{Q}(t)$ --- функция экспоненциально взвешенного
скользящего среднего (Exponentially Weighted Moving-Average,
EWMA)~\cite{floyd:1993:red}:
\begin{equation}
  \label{qhat}
\Hat{Q}(t)= (1-w_q)\Hat{Q}(t)+w_q Q(t),
\end{equation}
где $w_q$, $0< w_q< 1$ --- весовой коэффициент.

Мы использовали метод стохастизации одношаговых процессов
для получения уравнений
Фоккера--Планка и Ланжевена для случайных процессов $W(t)$ и $Q(t)$.

Кинетические уравнения будут иметь вид:
\begin{equation}
  \label{eq:kinetik:W:Q}
  \begin{dcases}
    0 \xrightarrow{\frac{1}{W(t)}} W (t), \\
    W (t) \xrightarrow{\frac{1}{2} \frac{\dd N(t)}{\dd t}} 0, \\
    0 \xrightarrow{\frac{W(t)}{T}} Q (t),\\
    0\xrightarrow{ - C} Q(t),
  \end{dcases}
\end{equation}
где $\dd N(t)$~--- пуассоновский процесс~\cite{misra:2000:fluid-based}.

Выпишем уравнения Фоккера--Планка, соответствующее кинетическим
уравнениям~\eqref{eq:kinetik:W:Q}:
\begin{equation}
  \label{eq:FP:W:Q}
  \begin{gathered}
    \begin{multlined}
  \frac{\partial w(t)}{\partial t} = 
  - \frac{\partial}{\partial W(t)} 
  \left[ \left( \frac{1}{W(t)} - \frac{W(t)}{2} \frac{\dd N(t)}{\dd t}
    \right) w(t) \right]
  + {} \\ {} 
  +
  \frac{1}{2} \frac{\partial^2}{\partial W^2(t)} 
  \left[ \left( \frac{1}{W(t)} + \frac{W(t)}{2} \frac{\dd N(t)}{\dd t}
    \right) w(t) \right],
\end{multlined}
\\
\begin{multlined}
  \frac{\partial q(t)}{\partial t} = - \frac{\partial}{\partial
    Q(t)}
  \left[ \left( \frac{W(t)}{T} - C \right) q(t) \right]
  + {} \\ {}
  +
  \frac{1}{2}
  \frac{\partial^2}{\partial Q^2(t)} \left[ \left( \frac{W(t)}{T} -
      C \right) q(t) \right],
\end{multlined}
\end{gathered}
\end{equation}
где $w(t)$~--- плотность распределения случайного процесса $W(t)$,
$q(t)$~--- плотность распределения случайного процесса $Q(t)$.

Соответствующие уравнениям~\eqref{eq:FP:W:Q} уравнения Ланжевена имеет
вид:
\begin{equation}
  \label{eq:W:Q}
  \begin{dcases}
      \dd W(t) = \frac{1}{T} \dd t - \frac{W(t)}{2} \dd N(t)
      +
      \sqrt{\frac{1}{T} + \frac{W(t)}{2} \frac{\dd N(t)}{\dd t}} \dd
      V^1(t),
    \\
    \dd Q(t)  = \left( \frac{W(t)}{T} - C \right) \dd Q(t) + \sqrt{\frac{W(t)}{T}
      - C}\dd V^2(t) ,
  \end{dcases}
\end{equation}
где $\dd V^1 (t)$~--- винеровский процесс, соответствующий случайному
процессу $W(t)$,
$\dd V^2(t)$~--- винеровский процесс, соответствующий случайному
процессу $Q(t)$.

Уравнения~\eqref{eq:W:Q}
дополняются уравнением связи (для удобства записанное в
дифференциальном виде):
\begin{equation}
  \label{eq:Qhat}
  \frac{\dd \Hat{Q}(t)}{\dd t}  = w_q C ( Q(t) - \Hat{Q}(t) ).
\end{equation}

Исследовать данную математическую модель можно как в виде
стохастических дифференциальных уравнений, так и записав их в
моментах. Уравнение в моментах, естественно, легче для исследования.

\section{Физическая модель}
\label{sec:physical-model}

Полученную математическую модель нужно сопоставить с
экспериментальными данными, верифицировать её. К сожалению, мы не
обладаем ресурсами для снятия показаний с работающей сети или для
построения натурного стенда на настоящем сетевом оборудовании. Поэтому
нами была произведена попутка собрать виртуальную экспериментальную
установку на основе виртуальных машин~\cite{kulyabov:2014:icumt-2014:gns3}.
На виртуальных машинах выполняются образы операционных систем реальных
маршрутизаторов. Именно это позволяет нам называть данную модель \emph{физической}.

Для построения стенда был выбран программный комплекс GNS3 (Graphical
Network Simulator)~\cite{welsh:2013:gns3}.  Он позволяет моделировать
виртуальную сеть из маршрутизаторов и виртуальных машин. Работает практически на всех
платформах. Фактически это графический интерфейс для разных
виртуальных машин. Для эмуляции устройств Cisco используется эмулятор
dynamips. Кроме того, можно использовать такие эмуляторы, как
VirtualBox и Qemu. Последний особенно удобен при использовании с
системой KVM, позволяющей использовать аппаратную реализацию
процессора. Фактически, GNS3 осуществляет оркестрацию работы разных
виртуальных машин, а также предоставляет исследователю удобный
интерфейс создания и настройки необходимой конфигурации стенда.
Кроме того, есть возможность соединения проектируемой
топологии с внешней сетью для управления и манипуляций пакетами данных.

Стенд состоит из маршрутизатора Cisco, генератора
трафика и получателя. В качестве генератора трафика используется D-ITG
(Distributed Internet Traffic Generator) (рис.~\ref{fig:stand}).
D-ITG позволяет получить оценки основных показателей качества
обслуживания (средняя задержка передачи пакетов, вариация задержки
(джиттер), коэффициент потерь пакетов, производительность) с высокой
степенью достоверности.

\begin{figure}
  \centering
  \includegraphics[width=0.6\linewidth]{virtual_network}
  \caption{Виртуальный стенд для исследования функционирования
    алгоритма RED. \texttt{host01}~--- источник пакетов; \texttt{host02}~--- получатель.}
  \label{fig:stand}
\end{figure}

\section{Имитационная модель}
\label{sec:simulation}

С развитием компьютерной техники появилась возможность задавать
модельную реализацию не в виде математического описания, а в виде
некоторого алгоритма (рис.~\ref{fig:model-simulation}). Такого типа
модели получили название имитационных моделей, а сам подход получил
название имитационного моделирования.

\begin{figure}
  \centering
  \includegraphics[width=0.6\linewidth]{model-simulation}
  \caption{Общая структура имитационной модели}
  \label{fig:model-simulation}
\end{figure}

Имитационная модель играет двоякую роль. Отлаженная и проверенная на
экспериментальных данных и физической модели имитационная модель сама
по себе может служить целям верификации математической модели.
С другой стороны, имитационная модель позволяет более эффективно, чем
математическая модель, исследовать поведение моделируемой системы при
разных вариантах входных данных.

\subsection{Имитационная модель на NS-2}
\label{sec:red:ns2}

Пакет ns2~\cite{issariyakul:2012,altman:2011} является средством
имитационного моделирования сетевых протоколов. За время своего
существования функционал многократно верифицировался данными натурных
экспериментов. Поэтому этот пакет сам стал эталонном средством
моделирования. Это именно тот случай, когда имитационная модель
является заменой физической модели и натурного эксперимента.

Программа для ns2 пишется на языке
TCL~\cite{welch:book:practical-tc-tk,nadkarni:book:tcl-comprehensive}.
Результаты моделирования можно представить с помощью 
средства визуализации nam
(см. рис.~\ref{fig:nam-drop}). 

Средство моделирования построено по событийно-ориентированной
архитектуре. То есть реализует дискретный подход к моделирования. С
одной стороны это является плюсом, поскольку напрямую реализует
спецификацию TCP и RED (см. раздел~\ref{sec:red-aqm}). С другой
стороны резко возрастает количество результирующих данных, что делает
затруднённым проведение сколь-либо длительного имитационного эксперимента.

\begin{figure}
  \centering
  \includegraphics[width=0.7\linewidth]{nam-drop}
  \caption{Визуализация моделирования системы. Показан сброс пакетов}
  \label{fig:nam-drop}
\end{figure}

В наших работах данный программный продукт используется именно как
средство верификации полученных результатов~\cite{kulyabov:2018:icumt:aqm,kulyabov:2019:aisc:self-oscillation}.

\subsection{Гибридная модель алгоритма RED}
\label{sec:red:model_hybrid}

Для исследования алгоритма RED нами был разработан прототип имитационной
модели. Мы хотели избежать ресурсозатратности дискретных подходов
моделирования. Но при этом необходимо было учесть дискретную
спецификацию TCP и RED (см. раздел~\ref{sec:red-aqm}). Поэтому нами
был выбран гибридный (непрерывно--дискретный) подход. Модель была
реализована на языке гибридного моделирования Modelica~\cite{fritzson:2003,fritzson:2011}.

Поскольку мы строим гибридную непрерывно--дискретную модель, то для
описания каждой фазы функционирования TCP перейдём к модели с
непрерывным временем. Переход же между фазами будет описываться
дискретными состояниями.

Для построения гибридной модели необходимо:
\begin{itemize}
\item записать динамическую модель для каждого состояния;
\item заменить системы с кусочно--постоянными параметрами на системы с
  переменными начальными условиями;
\item записать диаграмму состояний модели (рис.~\ref{fig:tcp:general}
  и~\ref{fig:red:general}).
\end{itemize}

Полученные диаграммы непосредственно
преобразовываются в программу на языке
Modelica~\cite{kulyabov:2017:iop:tcp-modelica,kulyabov:2019:ceur-ws:2407:red,kulyabov:2016:ecms:red}.

\begin{figure}
\begin{minipage}[b]{0.6\linewidth}
  \centering
  \includegraphics[width=\linewidth]{tcp_general}
  \caption{Диаграмма состояний TCP}
  \label{fig:tcp:general}
\end{minipage}
\hfill
\begin{minipage}[b]{0.35\linewidth}
  \centering
  \includegraphics[width=\linewidth]{red_general}
  \caption{Диаграмма состояний RED}
  \label{fig:red:general}
\end{minipage}
\end{figure}

\section{Суррогатная модель}
\label{sec:surrogate}

\begin{figure}
  \centering
  \includegraphics[width=0.6\linewidth]{model-surrogate}
  \caption{Общая структура суррогатной модели}
  \label{fig:model-surrogate}
\end{figure}

Большинство научных и технических проблем требуют
экспериментов и моделирования для получения результатов, определения
ограничений, накладываемых на результат.
Однако для многих реальных проблем одно только
моделирование может занять несколько минут, часов, дней. В
результате рутинные задачи, такие как оптимизация решений,
исследование пространства решений, анализ чувствительности и
анализ «что, если» становятся невозможными, поскольку они требуют
тысяч или миллионов оценок моделирования.

Один из способов упростить исследование~--- построить суррогатные
модели (аппроксимационные модели, модели поверхности отклика,
метамодели, модели чёрного ящика)
(см. рис.~\ref{fig:model-surrogate}), которые имитируют поведение
исходной модели настолько близко, насколько это возможно, в то время
как вычислительно
дёшевы~\cite{jin:2011:surrogate-assisted-computation}. Суррогатные
модели строятся с использованием подхода, основанного на
данных. Точная внутренняя работа кода моделирования не предполагается
известной (или даже понятой), важно только поведение ввода--вывода
(приготовления--измерения). Модель строится на основе моделирования
реакции на ограниченное количество (порой достаточно большое)
выбранных точек данных\footnote{Заметим, что такого типа модели
  известны многим исследователям. Когда задействована только одна
  расчётная переменная, процесс построения суррогатной модели
  называется подгонкой кривой}.

Научная задача суррогатного моделирования заключается в создании
суррогата, который является максимально точным, используя как можно
меньше оценок моделирования. Процесс состоит из следующих основных этапов,
которые могут чередоваться~\cite{kulyabov:2019:ceur-ws:2507:deep-learning}:
\begin{itemize}
\item выбор образца;
\item построение суррогатной модели и оптимизация параметров модели;
\item оценка точности суррогата.
\end{itemize}

Для некоторых проблем природа истинной функции априори неизвестна,
поэтому неясно, какая суррогатная модель будет наиболее точной. Кроме
того, непонятно, как получить наиболее надёжные оценки
точности данного суррогата. В данном случае модельный слой
(рис.~\ref{fig:model-surrogate}) заменяется догадками исследователя.
В нашем же случае суррогатная модель базируется на ясно
сформулированной математической модели, что позволяет получить ясные,
обоснованные результаты суррогатного моделирования.

На данный момент нами разрабатывается методика построения суррогатных
моделей как для собственно алгоритмов типа RED, так и для произвольных
стохастических одношаговых процессов.

\section{Заключение}
\label{sec:conclusion}

Авторы попытались обрисовать концепцию мультимодельного подхода к
исследованию физических и технических систем на примере
взаимодействующих протокола TCP и алгоритма активного управления
трафиком типа RED. Данное исследование находится в русле исследований
стохастических моделей в науке и технике.

Мультимодельный подход позволяет повысить эффективность исследование
явления, рассмотреть его с разных сторон и создать эффективные программные
комплексы.

\def\acknowledgmentsname{Благодарности}

\begin{acknowledgments}

Публикация подготовлена при поддержке Программы РУДН «5-100»
и при финансовой поддержке РФФИ в рамках научного проекта
\textnumero~19-01-00645.

\end{acknowledgments}

  \ifthenelse{\boolean{xetex} \OR \boolean{luatex}}
  {\bibliographystyle{ugost2008l}}
  {\bibliographystyle{gost2008l}}

\bibliography{bib/red-model-approach/cite}